\documentclass[nohyperref]{article}

\usepackage{microtype}
\usepackage{graphicx}
\usepackage{subfigure}
\usepackage{pifont}
\usepackage{booktabs} % for professional tables

\usepackage{hyperref}

\usepackage[accepted]{StyleFiles/icml2022}

\usepackage{amsmath}
\usepackage{amssymb}
\usepackage{mathtools}
\usepackage{framed}
\usepackage{amsthm}
\PassOptionsToPackage{hyphens}{url}\usepackage{hyperref}
\usepackage{xurl}
\usepackage[capitalize,noabbrev]{cleveref}

\theoremstyle{plain}

\theoremstyle{definition}

\theoremstyle{remark}

\usepackage[textsize=tiny]{todonotes}
\usepackage[font=small,skip=1pt]{caption}
\usepackage[resetlabels,labeled]{multibib}
\newcites{A}{References (Appendix)}

\makeatletter
\newcommand{\pushright}[1]{\ifmeasuring@#1\else\omit\hfill$\displaystyle#1$\fi\ignorespaces}
\newcommand{\pushleft}[1]{\ifmeasuring@#1\else\omit$\displaystyle#1$\hfill\fi\ignorespaces}
\makeatother

\makeatletter
\newcommand{\lrarrow}{\mathrel{\mathpalette\lrarrow@\relax}}
\newcommand{\lrarrow@}[2]{%
  \vcenter{\hbox{\ooalign{%
    $\m@th#1\mkern6mu\rightarrow$\cr
    \noalign{\vskip1pt}
    $\m@th#1\leftarrow\mkern6mu$\cr
  }}}%
}
\makeatother

\newcommand{\y}{\mathbf{y}}

\newcommand{\R}{\mathbb{R}}
\newcommand{\N}{\mathbb{N}}

\icmltitlerunning{N-ACT: Neural-Attention for Cell-Type Identification}

\begin{document}

\twocolumn[
\icmltitle{N-ACT: An Interpretable Deep Learning Model for Automatic Cell Type \\ and Salient Gene Identification}


\icmlsetsymbol{equal}{$\ddagger$}

\begin{icmlauthorlist}
\icmlauthor{A. Ali Heydari}{math,health,equal}
\icmlauthor{Oscar A. Davalos}{health,qsb,equal}
\icmlauthor{Katrina K. Hoyer}{health,mcb}
\icmlauthor{Suzanne S. Sindi}{math,health}
\end{icmlauthorlist}

\icmlaffiliation{math}{Department of Applied Mathematics, University of California}
\icmlaffiliation{qsb}{Quantitative and Systems Biology Graduate Program}
\icmlaffiliation{mcb}{Department of Molecular and Cell Biology, University of California, Merced, USA}
\icmlaffiliation{health}{Health Sciences Research Institute}

\icmlcorrespondingauthor{Suzanne Sindi}{ssindi@ucmerced.edu}

\icmlkeywords{Interpretable Deep Learning, Deep Learning, single-cell RNA sequencing, Attention, Machine Learning, Computational Biology, ICML}

\vskip 0.3in
]



\printAffiliationsAndNotice{\icmlEqualContribution} % otherwise use the standard text.

\begin{abstract}
Single-cell RNA sequencing (scRNAseq) is rapidly advancing our understanding of cellular composition within complex tissues and organisms. A major limitation in most scRNAseq analysis pipelines is the reliance on manual annotations to determine cell identities, which are time consuming, subjective, and require expertise. Given the surge in cell sequencing, supervised methods--especially deep learning models--have been developed for automatic cell type identification (ACTI), which achieve high accuracy and scalability. However, all existing deep learning frameworks for ACTI lack interpretability and are used as “black-box” models. We present \emph{N-ACT} (Neural-Attention for Cell Type identification): the first-of-its-kind interpretable deep neural network for ACTI utilizing neural attention to detect salient genes for use in cell-types identification. We compare N-ACT to conventional annotation methods on two previously manually annotated data sets, demonstrating that N-ACT accurately identifies marker genes and cell types in an unsupervised manner, while performing comparably on multiple data sets to current state-of-the-art model in traditional supervised ACTI.
\end{abstract}
\vspace{-20pt}
\section{Introduction}
\label{intro}
Single-cell RNA-sequencing (scRNAseq) technologies allow for measuring transcriptome-wide gene expression at the single-cell level. In contrast to bulk-RNA sequencing, scRNAseq can elucidate dynamic expression patterns between different cellular populations, providing a tremendous advantage when studying organisms as well as delineating intra-population heterogeneities \cite{scOmicsReviewPaper}.

Accurate identification of cell types in scRNAseq studies remains a challenging and time-consuming task \cite{ReviewOfAutomaticCellTypeID}. Cell type annotation is often performed manually by experts -- a long, laborious, and subjective process \cite{ReviewOfAutomaticCellTypeID, TutorialOnSingleCellAnnotation}. To mitigate these challenges, researchers have developed automatic cell type identification (ACTI) pipelines, including deep learning (DL) models such as ACTINN \cite{ACTINN}, which learn from datasets that have \emph{already been annotated from the same (or similar) populations} (see \cite{ReviewOfAutomaticCellTypeID} for a review on ACTI). However, the supervision required for these approaches limits their utility for studies which lack prior knowledge of tissue- or sample-specific cell types. Such models can not provide the much-needed biological interpretability on the algorithm's decision-making needed to translate scRNAseq findings to inform experimental design. 

% translate scRNAseq findings from in-silico studies to inform experimental design. 

% yumarys random sentences for trying to help us out here 

In this work, we introduce \emph{Neural-Attention for Cell Type identification (N-ACT)}: An interpretable \emph{unsupervised} DL model that employs \emph{attention} to detect salient genes, and applies this information to identify specific cell-types. Our results on multiple datasets show that N-ACT accurately predicts preliminary annotations with no prior knowledge about the system, providing a valuable complementary framework to experimental studies and computational pipelines.

% \begin{table}[H]
% \caption{Classification accuracies for naive Bayes and flexible
% Bayes on various data sets.}
% \label{sample-table}
% \vskip 0.15in
% \begin{center}
% \begin{small}
% \begin{sc}
% \begin{tabular}{lcccr}
% \toprule
% Data set & Naive & Flexible & Better? \\
% \midrule
% Breast    & 95.9$\pm$ 0.2& 96.7$\pm$ 0.2& $\surd$ \\
% Cleveland & 83.3$\pm$ 0.6& 80.0$\pm$ 0.6& $\times$\\
% Glass2    & 61.9$\pm$ 1.4& 83.8$\pm$ 0.7& $\surd$ \\
% Credit    & 74.8$\pm$ 0.5& 78.3$\pm$ 0.6&         \\
% Horse     & 73.3$\pm$ 0.9& 69.7$\pm$ 1.0& $\times$\\
% Meta      & 67.1$\pm$ 0.6& 76.5$\pm$ 0.5& $\surd$ \\
% Pima      & 75.1$\pm$ 0.6& 73.9$\pm$ 0.5&         \\
% Vehicle   & 44.9$\pm$ 0.6& 61.5$\pm$ 0.4& $\surd$ \\
% \bottomrule
% \end{tabular}
% \end{sc}
% \end{small}
% \end{center}
% \vskip -0.1in
% \end{table}
\vspace{-5 pt}
\section{Methods and Approach}
\label{methods}
\begin{figure*}
    \centering
    \includegraphics[width=\textwidth]{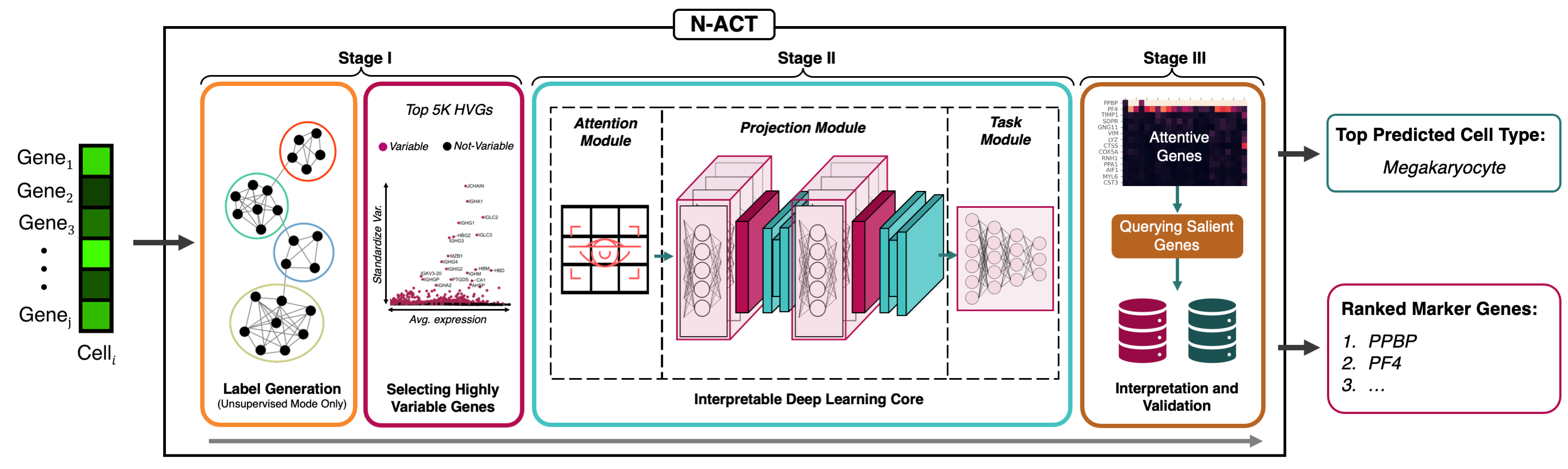}
    \caption{\textbf{N-ACT, an Interpretable Model for Unsupervised/Supervised Cell-Type Identification.} N-ACT consists of flexible stages and modules that can be modified for different objectives. We focus on unsupervised ACTI, which consists of the following stages: \emph{(Stage I)} Generating labels through graph-based clustering followed by identifying five thousand highly-variable genes for efficient training (standard practice in scRNAseq pipelines), \emph{(Stage II)} Training the N-ACT DL core to predict generated (or true) labels, with the goal of identifying salient genes to predict the correct cell types (labels), and \emph{(Stage III)} Interpreting model predictions by extracting attention values, thus constructing a ranked list of ``attentive genes" used to compare to existing literature (referred to as ``Querying Salient Genes" in the figure) and thus predicting cell types. We describe each component in more detail in the Appendix.}
    \label{fig:N-ACT}
\end{figure*}
% we're running into compiling issues my guy... can you compile on your end? did not compile
N-ACT's framework consists of three stages (shown in Fig. \ref{fig:N-ACT}): (I) Assigning labels (if no labels are available) and identifying highly-variable genes (HVGs), (II) a deep neural network (NN), which is the core of N-ACT, and (III) interpretation. In this section, we present stage II, i.e. N-ACT's DL core, and later describe stages (I) and (III) in Section \ref{results}. N-ACT's DL core has three modules: (1) \textit{Additive attention module}, responsible for learning the importance of each gene; (2) \textit{Multi-headed projection module} tasked with learning a set of non-linear operators mapping attention outputs to the last layer, \textit{i.e.} (3) \textit{Task module}, which is designed to fulfill a specific downstream task.
\vspace{-0.15 in}
\subsection{Additive Attention} 
\vspace{-0.1 in}
Attention (or neural-attention) is a recent DL mechanism that has transformed computer vision and natural language processing research (refer to \cite{ASurveryOfAttentionMechanism} for a review on attention in DL). Attention networks aim to mimic the way humans understand ``context" in sentences or details in images by focusing on a subset of significant features for a given objective. The use of attention-based NN for scRNAseq analysis is still in its infancy, with only a few successful applications to date. To identify salient genes, we use an additive attention module \cite{AdditiveAttention} in a feed-forward NN (similar to \cite{FFNN-Attn}) aiming to learn the optimal weighting (\emph{importance}) of all genes for each cell, given a downstream task.

The first step in the DL core (attention) is used to calculate a gene-score matrix (weighted version of scRNAseq count matrix), representing expression data in later layers. These importance scores enable gene prominence quantification for the downstream task, allowing for interpretation of model decision making. Given a gene expression matrix $X\in \R^{C \times N}$, where $C$ and $N$ denote the number of cells and genes, respectively, we define the gene-score matrix $\Gamma$ and the attention weights $A$ as shown in Eq. \eqref{eq:attention}: 
\begin{equation}
    \Gamma = A \odot X, \text{where } A_{i,j} = \frac{e^{L_{i,j}}}{\sum_{j=1}^N e^{L_{i,j}}},
    \label{eq:attention}
\end{equation}
with $L=NN(X)$ denoting a linear NN\footnote{$NN: \R^{C \times N} \to \R^{C \times N}$ is a linear operator of the form $NN(X) = X W + B$, with input $X$ and biases $B \in \R^{C\times N}$ and weights $W \in \R^{N\times N}$.}. The learned operator $A$ is leveraged after training to identify important (or ``attentive") genes for interpretability.
\vspace{-0.1 in}
\subsection{Multi-Headed Projections}
\vspace{-0.1 in}
Our \emph{Projection} mechanisms are intermediate layers between the attention layer and the downstream task module. The goal in using projection modules is to strike a balance between model capacity and efficiency: Too much capacity (\textit{e.g.} too many non-linear layers) could lead to significant over-fitting, while insufficient capacity prevents the model from learning the correct representations. We design the projection blocks to be \emph{multi-headed} (consisting of $h \in \N$ separate linear operators), a concept shown by \cite{AttentionIsAllYouNeed} as effective in learning different representations. Such design allows efficient consideration of different gene subsets and improves model performance without the need for numerous non-linear layers. N-ACT consists of $k$ projection blocks, each consisting of $h$ heads. Outputs from each head is then concatenated and inputted to a point-wise feed-forward network (equivalent to $1\times1$ convolution layer) with a Rectified Linear Unit (ReLU) \cite{ReLU} that adds non-linearity. Through careful ablation studies, we found that $k=2$ projection blocks with $h=10$ heads provides the appropriate balance of accuracy and efficiency. Details on projection blocks and relevant ablation studies are provided in \ref{sec:appendix-architecture}. 
\vspace{-0.15 in}
\subsection{Task Module and Architecture Choices}
\vspace{-0.1 in}
The last stage of our DL core is the \emph{task module}, which can be adjusted based on desired objectives. Given our ACTI goal, we chose a non-linear mapping between the projection block's output and the labels (either provided [supervised] or generated in the earlier stages [unsupervised]).

Our task module connects the projections to the number of labels, followed by Leaky ReLU activation \cite{LeakyReLU} (depicted in \ref{sec:appendix-architecture}). N-ACT minimizes a standard cross entropy loss (\ref{sec:appendix-math}) using the Adam gradient-based optimizer \cite{Adam} at a learning rate $\alpha=10^{-4}$ for 50 epochs. Additional information on N-ACT's training scheme is provided in \ref{sec:appendix-architecture}.

% \begin{figure}[ht]
% \vskip 0.2in
% \begin{center}
% \centerline{\includegraphics[width=\columnwidth]{icml_numpapers}}
% \caption{Historical locations and number of accepted papers for International
% Machine Learning Conferences (ICML 1993 -- ICML 2008) and International
% Workshops on Machine Learning (ML 1988 -- ML 1992). At the time this figure was
% produced, the number of accepted papers for ICML 2008 was unknown and instead
% estimated.}
% \label{icml-historical}
% \end{center}
% \vskip -0.2in
% \end{figure}
\vspace{-0.15 in}
\section{Datasets Studied}
\vspace{-0.1 in}
\label{sec:data}
We tested the model's ability to learn cell-types on four datasets (two for supervised and two for unsupervised ACTI). Results are presented on different datasets for each learning setting to showcase N-ACT's versatility and utility for different systems and species (similar results were achieved on all datasets in all tasks). Data were minimally pre-processed (only quality-controlled) and divided roughly 85\%:15\% for training and testing using ``balanced split" (described in \cite{ACTIVA}). A brief description of each dataset is provided below, with more details on pre-processing and the data presented in \ref{sec:appendix-datasets}.

\textbf{Datasets for supervised training}: (1) \textbf{\textit{Mouse HDF}} (PubMed ID: 34548614) consists of scRNAseq of murine aortic cells of mice on a normal diet versus mice on a high-fat diet, resulting in 24K cells and 10 annotated populations after processing. (2) \textbf{\textit{Immune CSF}} (PubMed ID: 33382973) profiles single-cells from cerebrospinal fluid (CSF) of immune CSF, viral encephalitis, and non-inflammatory and autoimmune neurological disease. Cells were isolated from 31 patients, resulting in 80K cells (70K cells after processing) and 15 annotated populations.

\textbf{Datasets for unsupervised training}: (1) \textbf{\textit{COVID PBMC}} (PubMed ID: 33357411) profiles the transcriptional immune dysfunction triggered in moderate and severe COVID-19 patients using scRNAseq. Peripheral blood mononuclear cells (PBMC) were isolated from 20 patients and were sequenced resulting in 69K cells (64K cells after processing) with 9 cell populations. (2) \textbf{\textit{Immune cSCC}} (Pub Med ID: 32579974) consists of scRNAseq from healthy skin and cutaneous squamous cell carcinoma (cSCC) tumors. 10 patients with cSCC tumors had healthy skin and tumor cells sequenced, resulting in 48K cells (47K cells after processing) and 14 cell populations.

\vspace{-0.15 in}
\section{Results}
\label{results}
In this section, we provide the results of using N-ACT for datasets described in Section \ref{sec:data}. Standard evaluation tools were used to measure model performance with each metric detailed in \ref{sec:appendix-math}.
\vspace{-0.1 in}
\subsection{Supervised ACTI Performance}
\label{sec:results-superACTI}
\vspace{-0.05 in}
We benchmark N-ACT against the current state-of-the-art supervised model, ACTINN, though the goal of this work is to generate unsupervised annotations. To show the importance of our novel architecture in effective attention utilization, we added the same feed-forward attention module to ACTINN (denoted by ``ACTINN+ATTN"), which significantly hindered the model's performance (Table \ref{tab:supervised-ACTI}). Given the comparable performance of our model to the state-of-the-art DL algorithm, our results show that N-ACT can effectively learn supervised ACTI. Moreover, the poor performance of ACTINN + ATNN highlights the importance of appropriate architectures needed for effective use of attention for interpretability. We next consider unsupervised cell-type annotation on two previously annotated datasets.

% Moreover, the performance of ACTINN with Attention  an adequate architecture is needed for effective utilization of attention for interpretability. 

\begin{table}[t]
\caption{\textbf{Benchmarking N-ACT on Supervised ACTI}. Although the main goal of N-ACT is interpretable ACTI in an unsupervised manner, our model can be used in a supervised setting as well, while still providing biological interpretability. \textit{W-F1}: Weighted F1 score, \textit{NW-F1}: Non-Weighted F1 score, \textit{Interp}: Interpretability }
\label{tab:supervised-ACTI}
\vspace{-0.05in}
\begin{center}
\begin{small}
\begin{sc}
\begin{tabular}{lcccr}
\toprule
Model & W-F1 & NW-F1 & Interp? \\
\midrule
&  Mouse HDF & & \\
\midrule
ACTINN              & \textbf{0.9703} & 0.9677              & No    \\
ACTINN+Attn         & 0.8759          & 0.7438              & Yes   \\
N-ACT (Ours)        & 0.9681          &  \textbf{0.9712}    & Yes   \\
\midrule
&  Immune CSF & & \\
\midrule
ACTINN              & \textbf{0.9357} & 0.8898              & No    \\
ACTINN+Attn         & 0.7528          & 0.2413              & Yes   \\
N-ACT (Ours)        & 0.9285          & \textbf{0.8963}     & Yes   \\
\bottomrule
\end{tabular}
\end{sc}
\end{small}
\end{center}
\vskip -0.3in
\end{table}

\vspace{-0.15 in}
\subsection{Unsupervised ACTI Performance}
\label{sec:results-unsuperACTI}
\vspace{-0.05 in}
\begin{figure*}[t]
    \centering
    \includegraphics[width=\textwidth]{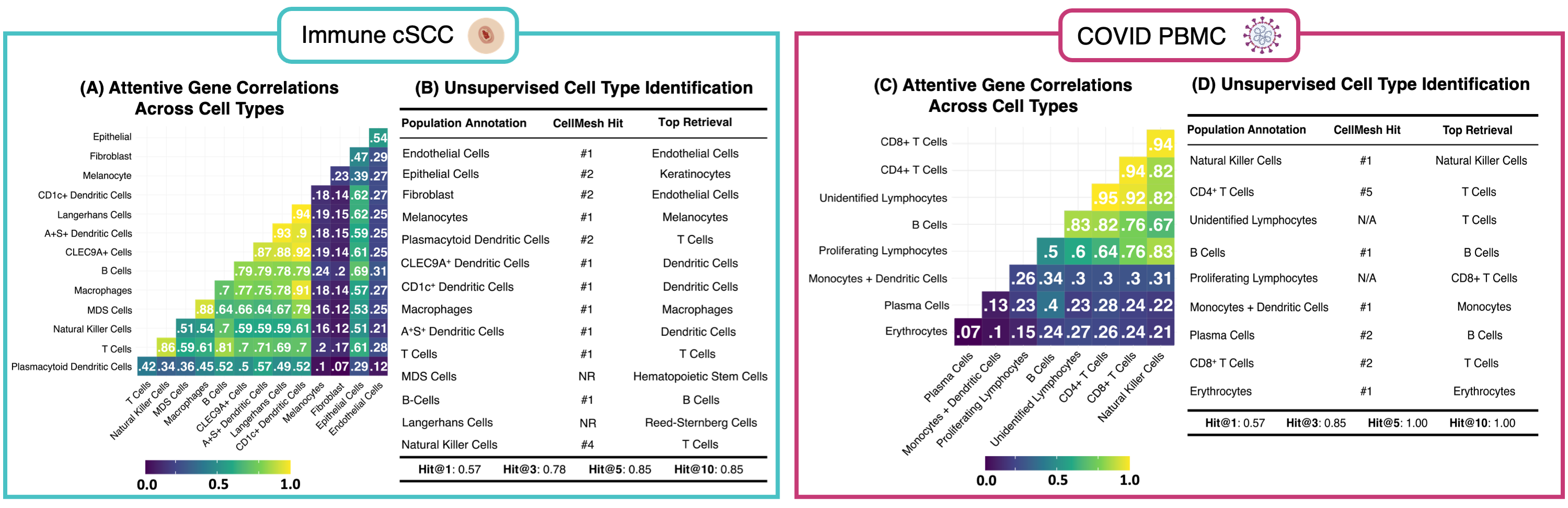}
    \caption{\textbf{Evaluating N-ACT's Unsupervised ACTI}. \textbf{[(A), (C)]:} To evaluate the interpretability and the performance of the attentive genes, we calculated the Pearson correlations of mean attention values in each population (per gene) for all cell types in \textbf{(A)} Immune cSCC and \textbf{(C)} COVID PBMC. \textbf{[(B), (D)]:} To assess N-ACT's predictive capabilities, we queried the TF-IDF-ranked attentive genes from CellMeSH and recorded the placement when the predicted cell type matched the actual annotation (called a “hit”). To interpret our model’s mistakes, we also retrieved the first prediction from CellMeSH, which we include as “Top Retrieval”.}
    \label{fig:Results}
\end{figure*}
\textbf{Unsupervised label generation}: Given that the cell types are not known \textit{a priori}, labels must be generated before training the DL core. To do so, we choose to perform unsupervised clustering using the Leiden algorithm \cite{Leiden}, a standard scRNAseq clustering technique in many pipelines \cite{DLinSpatialTranscriptomicsReview}, allowing label generation without supervision. All results shown in this section follow this approach. However, given the requisite comparison against the true annotations for this work, we ensure the number of clusters generated by our model is equal to the number of annotated populations from the original publication, without enforcing any additional constraints. 

\textbf{Finding salient genes}: To identify \emph{attentive} (salient) genes in each population, we leverage the attention scores calculated by our model: First, we calculate the mean attention score per cell type for each gene. To analyze the performance of our model in assigning importance to various genes, we investigate the correlation between the mean attention scores for all genes in each cell type (Fig. \ref{fig:Results}\textbf{(A),(C)}). We find a low correlation for those populations that are not closely related (\textit{e.g.} dendritic and endothelial cells in Fig. \ref{fig:Results}\textbf{(A)}), and a high correlation between those that are related (\textit{e.g.} CD4$^{+}$ and CD8$^{+}$ T cells in Fig. \ref{fig:Results}\textbf{(C)}). These results indicate that N-ACT has accurately learned to assign importance to the same gene sets across similar populations, as desired.

Next, we use mean attention scores per cluster and selected the top 100 genes with the highest average values, constructing an object that includes the top 100 genes for each cell per cluster. Using this object, we calculate term frequency (TF)-inverse document frequency (IDF) for each gene in the object. TF-IDF is a standard natural language processing technique that weights the importance of a word in a document (see \ref{sec:appendix-math} for more information). In this formulation, each row of the object (containing top genes for all cells in a cluster) is a document used to calculate TF and IDF. The TF-IDF values for each gene were then multiplied by the original attention values, providing the final saliency scores. TF-IDF normalization down-weights common housekeeping genes that frequently appear in each cell population but are less useful in identifying cell-types. Lastly, we re-rank genes based on these weighted scores and select the top 25 as the attentive genes, which we use for cell-type identification. 
 
\textbf{Identifying cell types}: Once attentive genes are identified, various techniques for finding their corresponding cell-types can be applied. To showcase the accuracy and automation capabilities of our model, we queried the attentive genes using CellMeSH \cite{CellMesh}, a probabilistic cell type querying tool that uses a database built from indexed literature to map marker genes to probable cell types (CellMeSH details provided in \ref{sec:appendix-CellMeSH}). It is important to note that the bias in each database can affect results, and that other specialized databases or methods can further improve the identification process (see \ref{sec:appendix-azimuth}). Fig. \ref{fig:Results}\textbf{(B)} and \ref{fig:Results}\textbf{(D)} present the accuracy of cell-type prediction using N-ACT-identified salient genes for Immune cSCC and for COVID PBMC, respectively. These results demonstrate that N-ACT accurately identifies attentive genes that are known markers for the underlying populations without any prior knowledge of the system or species.

%\oscar{be a cell type predictor. For this example, we leverage the attention scores generated for each gene in all clusters. First, we calculate the average attention score for each gene per cluster. Next, take the top 25 genes with the highest attention scores per cluster and create an object only containing gene names. Using the top 25 genes per cluster object, we calculate the term frequency-inverse document frequency (TF-IDF) for each gene in the object. TF-IDF is frequently applied in natural language processing to determine the importance of a word in a document. We consider each row in the object to be a document in which we calculate gene frequencies, and then we calculate the inverse document frequency, which down weights more common genes in the object. We multiply both term frequency and inverse document frequencies for each gene to obtain the tf-idf score. The generated tf-idf score down weights more common genes in the top 25 genes by multiplying the scores by the average attention scores. Once the attention scores have been weighted, the top 25 genes per cluster are submitted to CellMeSH (CITE: PMID: 34893819), which generates a prediction cell type. CellMesh is a probabilistic cell type prediction tool that uses a database built from the National Library of Medicine's (NLM) MEDLINE indexed records called Medical Subject Headings (MeSH). Explain CellMeSH probabilistic model here. 
\vspace{-0.15 in}
\section{Conclusions and Discussion}
\vspace{-0.05 in}
\label{conclusions}
In this work, we presented N-ACT, the first-of-its-kind interpretable DL model for ACTI. We show that N-ACT effectively identifies cell-types, in a supervised and, more importantly, unsupervised manner. N-ACT is a first attempt at providing interpretability in this context, and we believe our improvements and developments reduce subjectivity while significantly minimizing the time needed for annotating scRNAseq datasets. Optimizing the scRNAseq annotation process will accelerate translational and basic research by enabling scientists to focus on the underlying biological questions. Our results demonstrate that N-ACT accurately identifies salient genes that are known markers for the underlying populations, without prior knowledge of the system or species. Moreover, the interpretability of our framework is useful for predicting the correct cell-types, and for better understanding the data when there is ambiguity (\textit{e.g.} \ref{sec:appendix-azimuth}) or when the model makes mistakes. As such, N-ACT provides a powerful tool for facilitating discovery, even if its top prediction is ultimately incorrect. Using our model, cells can be assigned cell-types by new users without prior expertise in the given system, within minutes. From these conclusions, we hypothesize that attention can be further utilized to identify unique relationships between different genes and cells, which would not otherwise be apparent. Despite successful application of DL in scRNAseq space, most DL models are not interpretable. Biologically-interpretable DL models, such as N-ACT, can provide crucial information on the algorithm's decision making, while assisting scientists in understanding underlying complex biological networks.

\vspace{-0.15 in}
\section*{Code and Data Accessibility}
\vspace{-0.1 in}
All source code and reproducibility/tutorial notebooks, alongside download links to trained models and datasets, are available at \url{https://github.com/SindiLab/NACT}.
\clearpage
\vspace{-0.15 in}
\section*{Acknowledgements}
\vspace{-0.1 in}
The authors were supported from the National Institutes of Health (R15-HL146779 \& R01-GM126548), National Science Foundation (NSF) (DMS-1840265) and University of California (UC) Office of the President and UC Merced COVID-19 Seed Grant. Computational resources were supported by the NSF, Grant No. ACI-2019144 and in part through NSF awards CNS-1730158, ACI-1540112 and ACI-1541349. Due to space constraints, complementary citations are included in the Appendix.

% R15HL146779-01
% Diversity supplement A19-0028-002 


\bibliography{refs}
\bibliographystyle{StyleFiles/icml2022}

%%%%%%%%%%%%%%%%%%%%%%%%%%%%%%%%%%%%%%%%%%%%%%%%%%%%%%%%%%%%%%%%%%%%%%%%%%%%%%%
%%%%%%%%%%%%%%%%%%%%%%%%%%%%%%%%%%%%%%%%%%%%%%%%%%%%%%%%%%%%%%%%%%%%%%%%%%%%%%%
% APPENDIX
%%%%%%%%%%%%%%%%%%%%%%%%%%%%%%%%%%%%%%%%%%%%%%%%%%%%%%%%%%%%%%%%%%%%%%%%%%%%%%%
%%%%%%%%%%%%%%%%%%%%%%%%%%%%%%%%%%%%%%%%%%%%%%%%%%%%%%%%%%%%%%%%%%%%%%%%%%%%%%%
\newpage
\appendix
\onecolumn

\setcounter{page}{0}
    \pagenumbering{arabic}
    \setcounter{page}{1}

\setcounter{section}{0}
\renewcommand{\thesection}{Appendix \Alph{section}} 
\renewcommand{\thesubsection}{\Alph{section}. \arabic{subsection}}
\onecolumn
\noindent
{\huge \textbf{Appendix and Supplementary Material}}\\[1cm]
%\beginsupplement

\renewcommand{\theequation}{A.\arabic{equation}}
\renewcommand{\thefigure}{A.\arabic{figure}}
\renewcommand{\thetable}{A.\arabic{table}}

\section{Computational Environment}
\label{sec:appendix-DevEnv}
Model development and testing was performed in \texttt{Python} (v 3.9.7) and data processing was performed in \texttt{R} (v. 4.1.2) (detailed in section B). Our models were developed and tested on A100 Nvidia GPUs, and data pre-processing on a 14-inch Macbook Pro with Apple M1 Max and 64 GB RAM. Data I/O was done in \texttt{Scanpy} (v. 1.7.0) \citeA{scanpy}. The DL core of our model was developed in \texttt{PyTorch} (v. 1.9.1); however, developing/testing N-ACT on A100 GPUs required installation of a specific version of \texttt{PyTorch} (\texttt{torch==1.9.0+cu111}), which is provided in N-ACT's package repository. A complete list of requirements of \texttt{Python} packages is also available in N-ACT's GitHub repository: \url{https://github.com/SindiLab/NACT}.

\section{ Datasets Studied and Pre-Processing Workflow}
\label{sec:appendix-datasets}
\subsection{Pre-Processing}
All data (count matrices and manual annotations) are publicly available from NCBI gene expression omnibus (GEO) and the Broad Institute Single Cell Portal (SCP), with links provided below. Datasets were processed using the \texttt{Seurat} package (v. 4.1.0) in \texttt{R} \citeA{Seurat}. Manual annotations were merged with count matrices using a variety of \texttt{tidyverse} (v. 1.3.1) functions, and subsequently added to \texttt{Seurat} object as metadata. Data filtering consisted of the standard practice of removing cells with fewer than 200 expressed genes and removing genes present in fewer than 3 cells. Next, we retained cells with less than 10\% mitochondrial reads to mitigate cellular debris. Lastly, cell types containing less than 100 cells were removed and excluded from the dataset.

After filtering, we identified the top 5,000 highly variable genes (HVG's) for each dataset (HVG procedure is detailed in Appendix \ref{sec:HVG}). To minimize the biological and technical effects in each dataset based on patient and/or biological conditions such as normal versus disease state, we utilized \texttt{Harmony} (v 0.1.0) to perform integration \citeA{Harmony}. Resulting Uniform Manifold Approximation and Projection (UMAP) \citeA{UMAP} plots for each dataset with original annotations are provided in Fig. \ref{fig:appendix-datasets}. \texttt{SeuratDisk} (v. 0.0.0.9019) was used to convert the Seurat data object into an AnnData object compatible with \texttt{Scanpy}. To perform clustering and generate cell labels (in the unsupervised case), we used \texttt{Scanpy}'s pipeline for clustering (consisting of dimensionality reduction using principal component analysis (PCA), followed by Leiden clustering). As mentioned in the main manuscript, we found Leiden resolutions that led to the same number of clusters as the annotated populations in order to compare our predictions to the ground truth labels.

\subsection{Datasets}
\begin{figure}[H]
    \centering
    \includegraphics[width=0.9\textwidth]{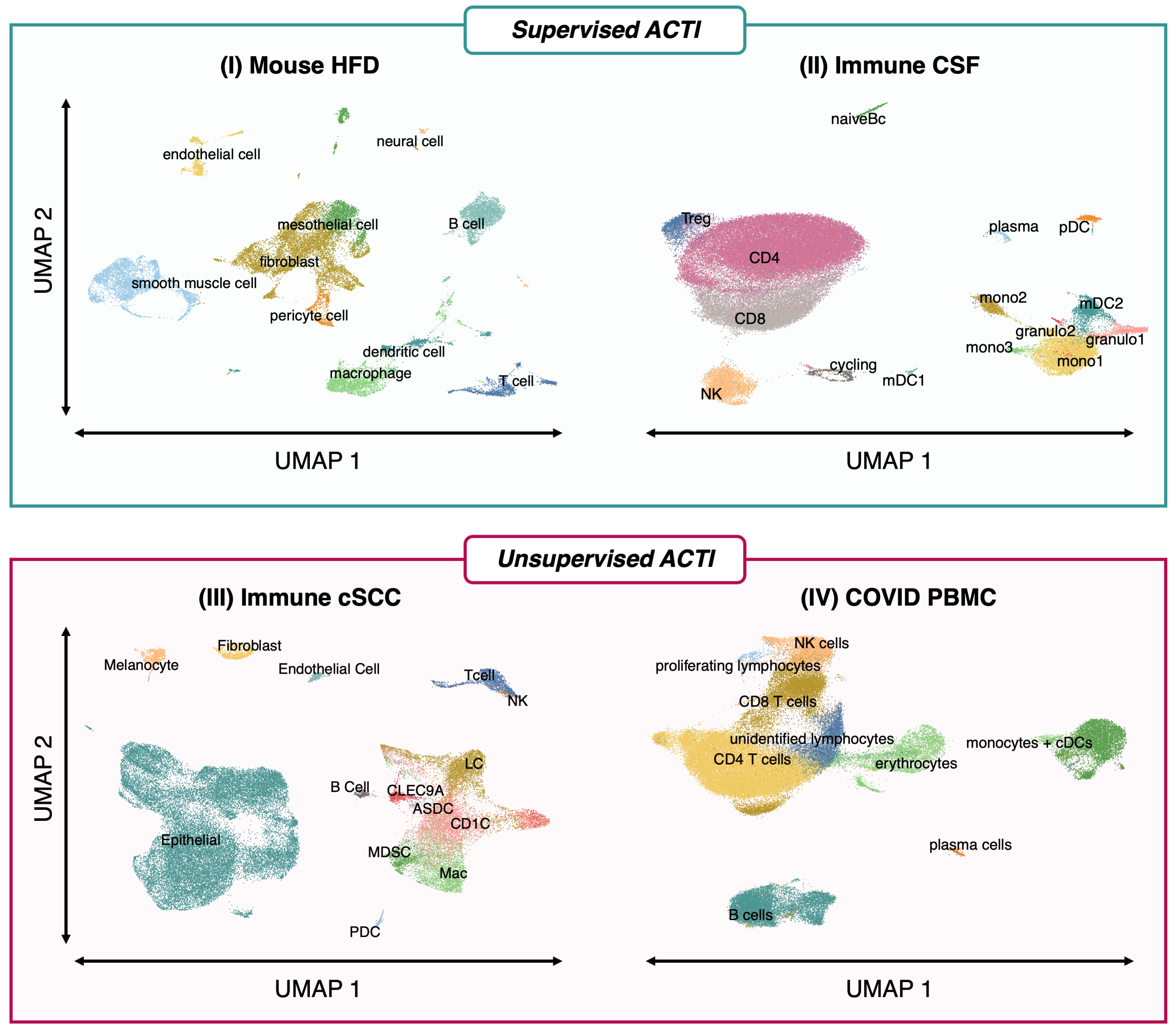}
    \caption{A UMAP-reduced plot of cell populations present in each dataset investigated in this work.}
    \label{fig:appendix-datasets}
\end{figure}

To evaluate N-ACT capabilities, we chose four large datasets relevant to immune researchers in light of the current pandemic. Three datasets are comprised of human cells, and one dataset consists of mouse cells. The three human datasets were chosen due to the distinct disease conditions being evaluated in the original studies. These included two COVID viral infection studies and human cSCC cancer study. We included a mouse dataset to demonstrate our model can be effectively used on non-human and non-immune datasets. All datasets were generated using the 10x Genomics platform (see \citeA{10XplatformReview, DLinSpatialTranscriptomicsReview-App} for a review of next-generation scRNAseq). A summary for each dataset is provided in subsequent subsections (see Fig. \ref{fig:appendix-datasets}). 
\subsubsection{Mouse-HDF}
Mouse-HDF \citeA{MouseHDF} [SCP1361] consists of scRNAseq of aortic cells in mice given a normal diet and mice given a high-fat diet (HFD), resulting in 24K cells (24K cells after processing). The authors identified 27 clusters, for 10 different cell populations.

\textit{Download Link}: \url{https://singlecell.broadinstitute.org/single_cell/study/SCP1361/single-cell-transcriptome-analysis-reveals-cellular-heterogeneity-in-the-ascending-aorta-of-normal-and-high-fat-diet-mice}

\subsubsection{Immune-CSF}
Immune-CSF \citeA{ImmuneCSF} [GSE163005] profiles single cells in cerebrospinal fluid (CSF) of Neuro-COVID, non-inflammatory, autoimmune neurological diseases, and viral encephalitis patients. Cells isolated from 31 patients: 8 COVID patients, 9 non-inflammatory, 9 autoimmune, 5 viral encephalitis resulting in a total 80K cells. After processing the dataset contains 70K cells with 15 populations. 

\textit{Download Link}: \url{https://www.ncbi.nlm.nih.gov/geo/query/acc.cgi?acc=GSE163005}

\subsubsection{COVID PBMC}
COVID-PBMC \citeA{COVIDPBMC} [GSE163005] consists of the evaluation of the transcriptional immune dysfunction triggered during moderate and severe COVID-19 patients using scRNA-seq. Peripheral blood mononuclear cells (PBMC) were isolated from 20 patients and were sequenced. Patients included in the study ranged from healthy ($n=3$), moderate COVID ($n=5$), acute respiratory distress syndrome (ARDS-Severe, $n=6$), and recovering ARDS-Recovering ($n=6$), resulting in 69K cells. After pre-processing the dataset, we retained 64K cells. The authors identified 9 populations in the dataset.

\textit{Download Link}: \url{https://www.ncbi.nlm.nih.gov/geo/query/acc.cgi?acc=GSE154567}

\subsubsection{Immune-cSCC}
Immune-cSCC \citeA{ImmunecSCC} [GSE144236] evaluates scRNAseq of normal skin and patient cutaneous squamous cell carcinoma (cSCC) tumors. Patients (n=10) with the cSCC tumors had normal skin and tumor cells sequenced, resulting in 48K cells. We retained 47K cells after pre-processing.
The authors identified 7 major cell populations. The myeloid cell population (CD14+Hi) is composed of various subpopulations.

\textit{Download Link}: \url{https://www.ncbi.nlm.nih.gov/geo/query/acc.cgi?acc=GSE144236}

\section{: Definition of Standard Mathematical Expressions and Evaluation Metrics}
\label{sec:appendix-math}
\subsection{Cross Entropy Loss}
The downstream task was to predict the correct labels (original labels or generated labels), with the learning objective being a standard cross-entropy loss:
\begin{equation}
H(\y,\hat{\y})= \sum_{j=1}^k y_j \log(\hat{y}_j) + (1-y_j)\log(1-\hat{y}_j),
    \label{eq:appendix-crossEntropy}
\end{equation}
where $\y$ and $\hat{\y}$ denote the original/generated labels and model predicted labels, respectively. In this formulation, $\y$ is a one-hot encoded vector of the cell types; that is  
\begin{equation*}
y_j = 
\begin{cases}
1 \hspace{0.1 in} \text{if cell type is } j\\
0 \hspace{0.1 in} \text{otherwise.}
\end{cases}
\end{equation*}

\subsection{Formulation of TF-IDF for Cell Type Querying}
In this work, we leverage attention scores generated for each gene in all clusters. First, we calculate the average attention score for each gene per cluster. Next, take the top 100 genes with the highest attention scores per cluster and create an object containing only gene names. Using the top 100 genes per cluster object, we calculate the term frequency(TF)-inverse document frequency(IDF) for each gene in the matrix. TF-IDF is a standard tool in natural language processing (NLP) often used to weight the importance of words appearing in documents, as shown in \eqref{eq:appendix-tfidf}. 
% \begin{equation}
\begin{subequations}
\begin{align}
    &TF(g,P) = \frac{f_{P}(g)}{|P|} \\ \bigskip \bigskip
    &IDF(g,D) =-\log p(g|D) = \log \left(\frac{|D|}{| \{ P\in D, g \in P \} |} \right) \\ \bigskip \bigskip
    &TF\text{-}IDF(g,P,D) = TF(g,P) \cdot IDF(g,D) 
\end{align}
\label{eq:appendix-tfidf}
\end{subequations}

where $f_P$ is the raw count of term (gene) $g$ in the document (population) $P$, with a corpus of documents $D$. We consider each row in the matrix to be a document in which we calculate gene frequencies, and then calculate the IDF, which down-weights more common genes in the matrix. We multiply both TF and IDF for each gene to obtain the TF-IDF score. The generated TF-IDF score down-weights common genes in the top 100 by multiplying the TF-IDF scores by the average attention scores. Once the attention scores have been weighted, the top 25 genes per cluster are submitted to CellMeSH (CITE: PMID: 34893819), which generates a prediction cell type.

\subsection{Selecting Highly Variable Genes}
\label{sec:HVG}
To select Highly Variable Genes (HVGs), we utilized \texttt{Seurat}'s \texttt{FindVariableFeatures()} function, described in \citeA{HVG_V3}. In this step, the aim is to calculate a measure of single-cell dispersion while including the mean expression. Step one learns a mean-variance relationship from the data, which is then used to calculate an expected standard deviation for each gene. To do so, the mean $\mu_g$ and variance $\sigma_g$ of each gene is computed from raw data. Next, a curve fitting using a $2^{\text{nd}}$ degree polynomial is applied to learn $f(\mu_g) = \hat{\sigma}_g$, providing a regularized estimator of variance given the mean of a gene $g$. Using this, we perform feature transformations without removing higher-than-expected variations. That is, given the raw counts $X_{ig}$ for gene $g$ in cell $i$, the mean raw value for gene $g$, $\mu_g$, and $\sigma_g$, the expected standard deviation of gene $g$ from the global fit, we have the transformed count value of gene $g$ as:
\begin{equation}
    z_{ig} = \frac{X_{ig} - \mu_g}{\sigma_g},
\end{equation}
for all gene $g$ across all cell $i$. Finally, variance across the new standard feature values is calculated and used to rank the genes. Although the convention is to select 2,000 HVGs, we chose to select 5,000 HVGs to increase the complexity and truly test model interpretability. 

\subsection{Hit@k}
\begin{figure}[H]
    \centering
    \includegraphics[width=\textwidth]{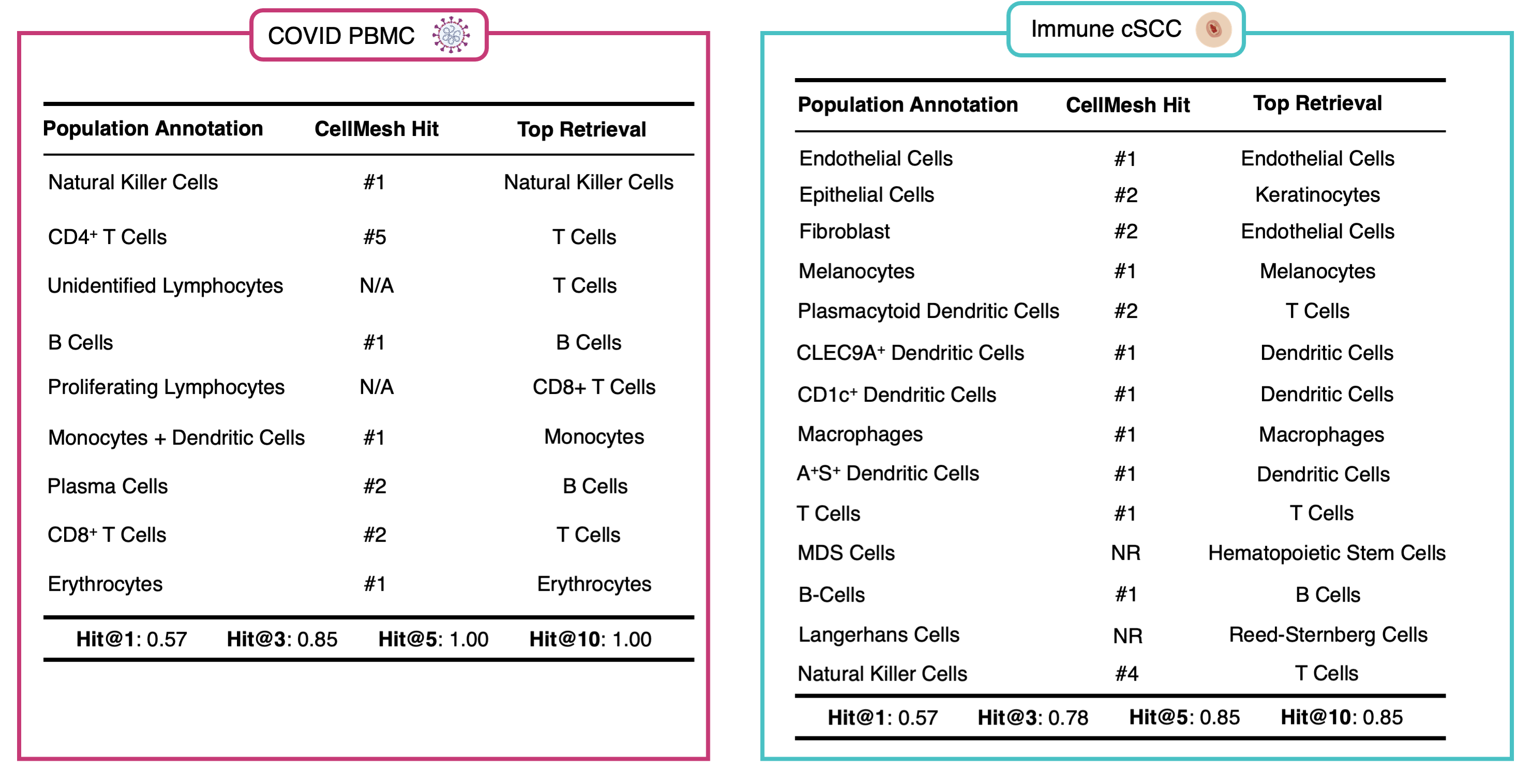}
    \caption{(For the ease of the reader) An enlarged version of Hit@K results, initially presented in Fig. \ref{fig:Results} of the main manuscript.}
    \label{fig:appendix-Hit@K}
\end{figure}

Cell type retrieval position is measured using N-ACT attentive genes for each cell population using \emph{Hit@k}. Hit@k is a metric measuring retrieval of a target value among the top $k$ retrievals. In our model, a ``hit" is the published cell type annotation (ground truth label) existing in the database. Rankings for $k=\{1,3,5,10\}$ are reported in Fig. \ref{fig:Results}, and in Fig. \ref{fig:appendix-Hit@K}. 

\subsection{F1 Score}
F1 score is a standard metric for evaluating a classifier. F1 score is the harmonic mean of precision and recall, which is shown in Eq. \eqref{eq:appendix-F1},
\begin{equation}
    F1 = 2\left(\frac{\text{precision}\cdot \text{recall}}{\text{precision} + \text{recall}}\right).
\label{eq:appendix-F1}
\end{equation}

For more information on weighted and non-weighted F1 score, see \url{https://scikit-learn.org/stable/modules/generated/sklearn.metrics.f1_score.html}.

\section{CellMeSH Cell Type Predictions}
\label{sec:appendix-CellMeSH}
CellMeSH \citeA{CellMesh} is a probabilistic method that generates a cell type prediction based on a gene list. CellMeSH aims to alleviate two common issues with querying gene lists from the literature: 1) publication bias, which relates to specific genes/cell types that are studied more often than other types, and therefore have more literature associated with them; and 2) Noise present in gene/cell-type mapping and the corresponding database. CellMesh uses a database built from the National Library of Medicine (NLM) MEDLINE indexed records called Medical Subject Headings. In building the CellMeSH database, Mao et al. use TF-IDF to reduce publication bias thus addressing publication bias. 

To address the second issue, CellMeSH utilizes a probabilistic querying method. Given gene query list $Q$, CellMeSH assumes a probabilistic model for a query genes $g$ being obtained from a cell-type $C$, shown in Eq. \eqref{eq:appendix-CellMesh}:
\begin{equation}
    P(g|C) = 
    \begin{cases}
    \alpha \cdot w_C(g) \hspace{0.2 in} &g \in Q \cap C \\
    (1-\alpha) \cdot \frac{1}{N_g - K_C} \hspace{0.2 in} &g \in Q \cap \overline{C}
    \end{cases}
    \label{eq:appendix-CellMesh}
\end{equation}
with $w_C(g)$ being the adjusted weight of gene $g$ in cell type $C$ (using TF-IDF), $N_g, K_C$ denoting the total number of genes and total number genes with non-zero weight in $C$. $\alpha$ is a parameter which aims to help with noise in the database. For each candidate cell type $C$, CellMeSH calculates a log-likelihood score:

\begin{equation}
    L(Q|C) = \log P(Q|C) = \sum_g \log P(g|C),
\end{equation}

which are then used to rank $C$ based on their values. Lastly, CellMeSH uses a maximum likelihood-based estimation to generate predictions. In this framework, the top cell type, $C^*$, can be found as:

\begin{equation}
    C^*= \text{argmax}_C \log L(Q|C).
\end{equation}

\section{: Main Manuscript Results (Larger Figures)}

\begin{figure}[H]
    \centering
    \includegraphics[width=\textwidth]{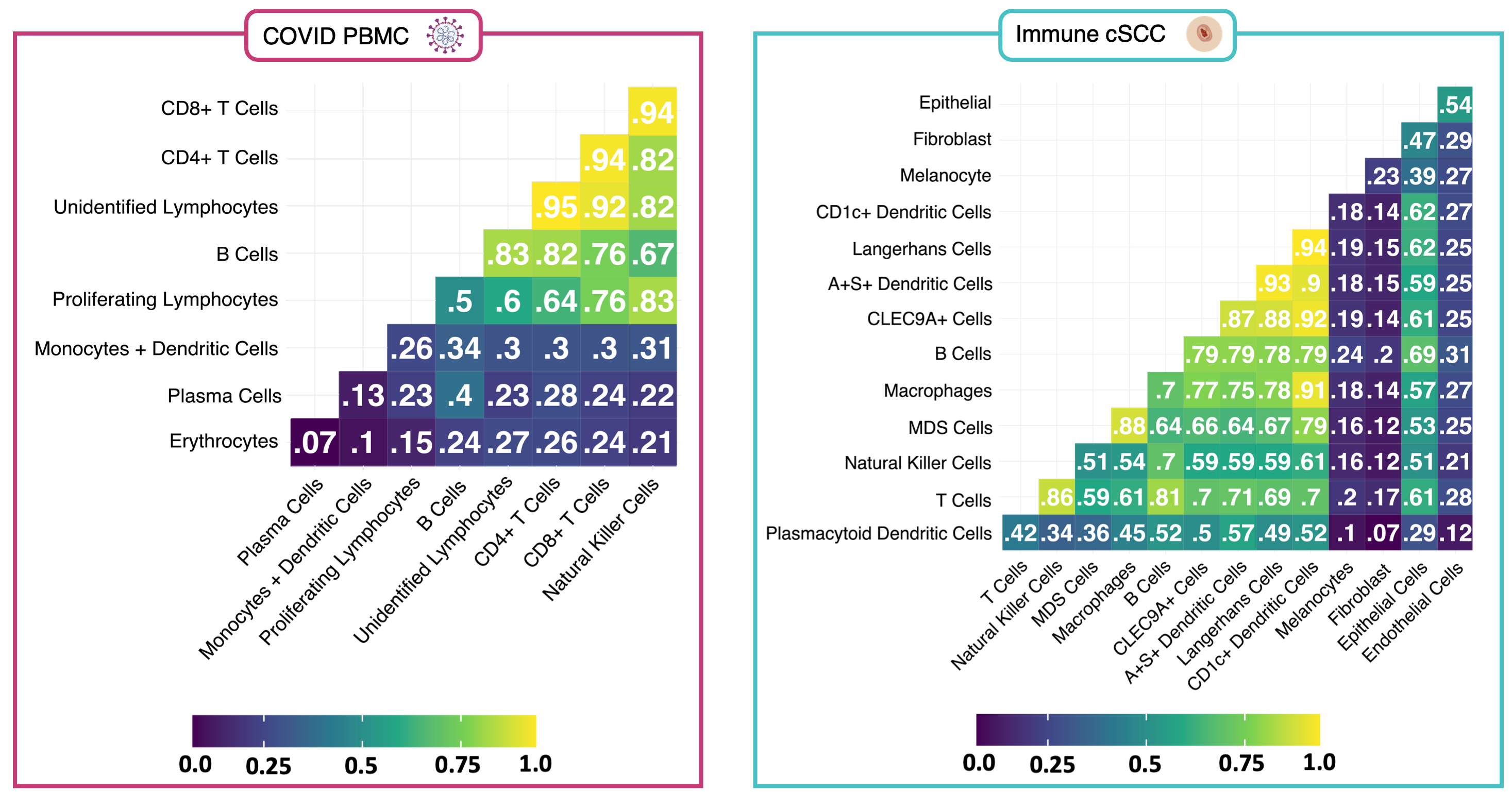}
    \caption{(For the ease of the reader) An enlarged version of correlation results presented in Fig. \ref{fig:Results} of the main manuscript.}
    \label{fig:appendix-Correlation}
\end{figure}

\section{N-ACT Architecture Details} 
\label{sec:appendix-architecture}

 \begin{figure}[H]
 \begin{framed}
    \centering
    \includegraphics[width=\textwidth]{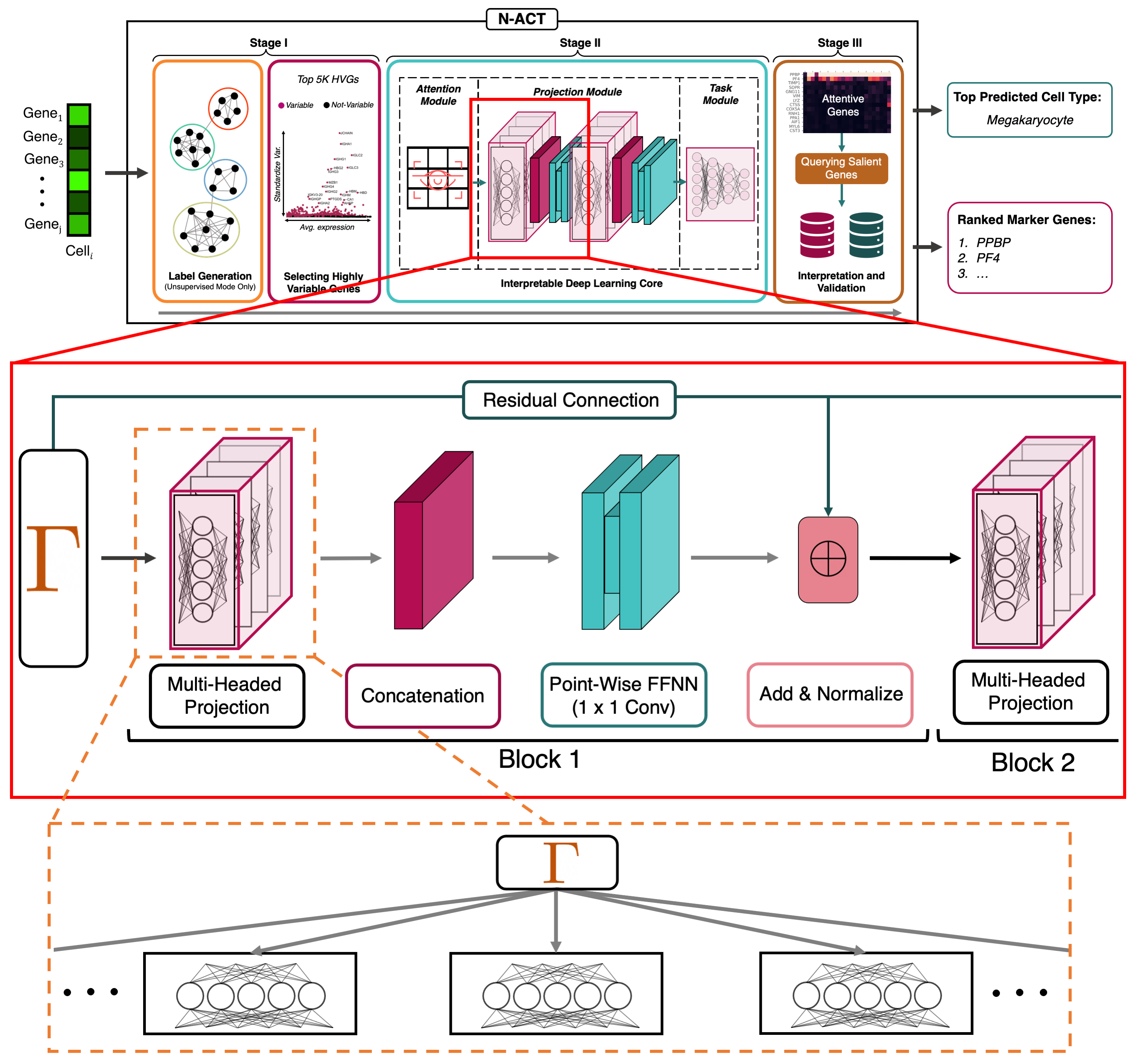}
    \smallskip
    \caption{\textbf{A Detailed View of N-ACT and the Multi-Headed Projection Module}. We dive deeper into the projection block of N-ACT in \ref{sec:appendix-architecture}. An implementation view of the DL architecture is presented in Fig. \ref{fig:appendix-implementationView}.}
    \label{fig:appendix-projection}
    \end{framed}
\end{figure}

\subsection{Projection Block}
The idea behind the N-ACT projection block is to learn various representations for different gene subsets in each cell. Projection block design was inspired by the multi-head attention architecture presented in \citeA{AttentionIsAllYouNeed-App}; however, the projections are not multi-head attention mechanisms. One way to think about the multi-head projection block is to view it as a set of $h$ linear projections, with each $l_h: \R^{B\times N} \to \R^{B\times d}$ ($B$ is the number of samples and $N$ is the number of genes) done sequentially and independent of one-another (\textit{e.g.} in a \texttt{for} loop). However, as noted by \citeA{AttentionIsAllYouNeed-App}, these projections can be done more efficiently through creating a tensor $L \in \R^{B\times h\times d}$, which acts the same way as the collection of the individual linear operators. In this formulation, $0 \equiv N (\text{mod } h)$, and the last projection component, the ``concatenation" layer (depicted in Fig. \ref{fig:appendix-projection}) allows us to reshape the model output which can be passed along to remaining layers. 

To increase model capacity and allow N-ACT to learn complex mappings, outputs are non-linearly ``activated" through a Point-Wise Feed Forward Neural Network (which can be thought of as 1$\times$ 1 convolution). However, there is the possibility of projection blocks learning a non-linear mapping unrelated to interpretability, resulting in loss of interpretability. Therefore, we hypothesized that adding residual connection between output of the attention module and input of each subsequent layer would improve performance and interpretability. Our ablation studies show that adding a residual connection improves model performance (Table \ref{tab:appendix-residual}). Skip connections are added to projection block outputs and normalized using Layer-Norm \citeA{LayerNorm}) before being used as input for the next layer. 

\begin{figure}
    \centering
    \includegraphics[width=\textwidth]{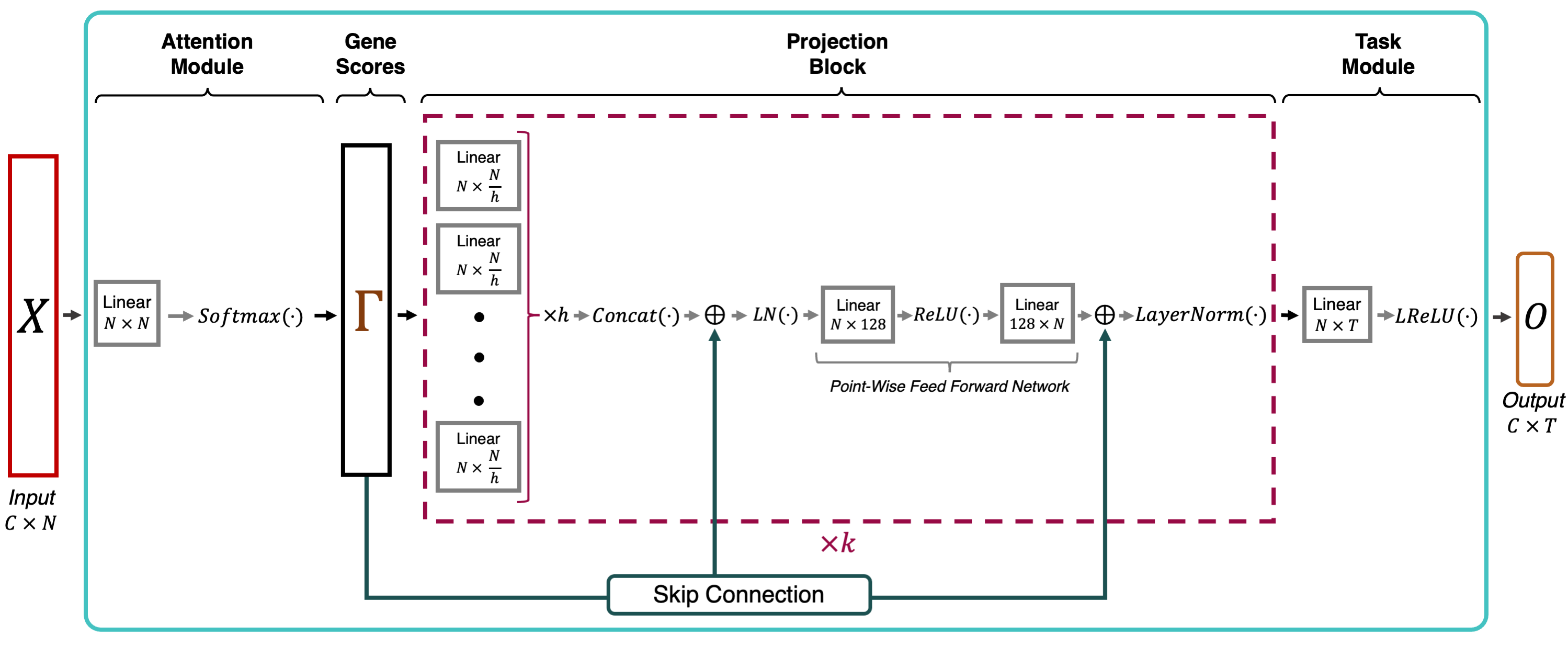}
    \caption{\textbf{An Implementation View of the N-ACT DL Core}. In this illustration, $C$ corresponds to the number of cells inputted, $N$ denotes the number of features ($N=5000$ in our results), $h$ is the number of projection heads, $k$ is the number of projection blocks(in our model, $h = 10$ and $k=2$) and $T$ denotes the number of classes (distinct labels) present in the data. }
    \label{fig:appendix-implementationView}
\end{figure}

\subsection{Multi-Head Ablation Study}
We investigated the effect of head number in the multi-head projection block and found that 10 heads provided the best results (Table \ref{tab:appendix-ablation}). We also tested our model accuracy when using one, two and three projection blocks and found that two projection blocks provided the best balance of accuracy and efficiency. To test our hypothesis regarding residual connections, we tested N-ACT (10 heads, 2 projection blocks) with and without skip connections. As shown in Table \ref{tab:appendix-residual}, model accuracy drops without the residual connections. Residual connections are identity mappings added to the output of each layer, with this quantity normalized and used as inputs for subsequent layers.
\begin{table}[H]
\caption{\textbf{Ablation Study on the Number of Projection Heads.} We fixed all other hyperparameters and studied the effect of head number on accuracy and interpretability. Training times are the average of 5 training runs on an A100 GPU. Accuracy of each model remained the same across different training settings, since all random parameters were initialized with the same random seed.}
\label{tab:appendix-ablation}
\vspace{-0.05in}
\begin{center}
\begin{small}
\begin{sc}
\begin{tabular}{lcccr}
\toprule
Number of Heads     & W-F1              & NW-F1             & Hit@5         & Avg. Training Time\\
\midrule

\hspace{0.5in}1                   & 0.9278            & 0.8968            & 0.85          & $9.11 \pm 0.14 \text{ (Min)}$  \\
\hspace{0.5in}5                   & 0.9317            & 0.9077            & 0.85          & $9.38 \pm 0.31 \text{ (Min)}$    \\
\hspace{0.5in}8                   & 0.9307            & 0.9156            & \textbf{1.00} & $9.31 \pm 0.22 \text{ (Min)}$   \\
\hspace{0.45in}10                  & 0.9322            & \textbf{0.9173}          & \textbf{1.00} & $9.56 \pm 0.27 \text{ (Min)}$ \\
\hspace{0.45in}20                  & \textbf{0.9324}            & 0.9156          & \textbf{1.00} & $9.87 \pm 0.24 \text{ (Min)}$ \\
\bottomrule
\end{tabular}
\end{sc}
\end{small}
\end{center}
\vskip -0.3in
\end{table}

\begin{table}[H]
\caption{\textbf{Effectiveness of Residual Connections on N-ACT.} }
\label{tab:appendix-residual}
\vspace{-0.05in}
\begin{center}
\begin{small}
\begin{sc}
\begin{tabular}{lcccr}
\toprule
Residual Connection?     & W-F1              & NW-F1     \\
\midrule
\hspace{0.55 in} Yes               & \textbf{0.9322}            & \textbf{0.9173}     \\
\hspace{0.6 in}No            & 0.9243            & 0.8725     \\
\bottomrule
\end{tabular}
\end{sc}
\end{small}
\end{center}
\vskip -0.3in
\end{table}

\subsection{Training Scheme}
As mentioned in the main manuscript, we train N-ACT for 50 epochs using Adam \citeA{Adam-App} optimizer, with a fixed learning rate. We tested training N-ACT for more epochs and found that the accuracy of predictions increases; however, a learning rate scheduling is beneficial in avoiding overfitting (when training over 100 epochs). When training for longer epochs, we employed an exponential learning rate decay with $\gamma=0.95$ and a decay schedule of every 10 epochs. We chose 50 epochs to balance accuracy and training time efficiency. 

\subsection{Supervised ACTI Comparison}

\begin{table}[H]
\caption{\textbf{Benchmarking N-ACT on Supervised ACTI}. Although the main goal of N-ACT is interpretable ACTI in an unsupervised manner, our model can be used in a supervised setting as well, while still providing biological interpretability. The reported training times are the median of 5 runs for each model. The same random seed was used to initialize all random parameters to ensure reproducibility of results across different runs. \textit{Training/Testing Support}: the number of samples used in training/evaluation, \textit{W-F1}: Weighted F1 score, \textit{NW-F1}: Non-Weighted F1 score}
\label{tab:appendix-supervised-ACTI}
\vspace{-0.05in}
\begin{center}
\begin{small}
\begin{sc}
\begin{tabular}{lccccr}
\toprule
Model               & W-F1            & NW-F1               & Training Support      & Testing Support       & Med. Training Time \\
\midrule
& &  & \textbf{Mouse HDF} & & \\
\midrule
ACTINN              & \textbf{0.9703} & 0.9677              & 21,003             & 2,998                      & \textbf{1.5 Minutes} \\
N-ACT (Ours)        & 0.9681          &  \textbf{0.9712}    & 21,003             & 2,998                      &  5.4 Minutes         \\
\midrule
 & & & \textbf{Immune CSF} & & \\
\midrule
ACTINN              & \textbf{0.9357} & 0.8898              & 66,549             & 8,991                      & \textbf{3.8 Minutes} \\
N-ACT (Ours)        & 0.9285          & \textbf{0.8963}     & 66,549             & 8,991                      &  11.2 Minutes         \\
\midrule
 & & & \textbf{COVID PBMC} & & \\
\midrule
ACTINN              & \textbf{0.9323} & 0.9148              & 55,005             & 9,864                      & \textbf{3.1 Minutes} \\
N-ACT (Ours)        & 0.9322          & \textbf{0.9173}     & 55,005             & 9,864                     & 9.4 Minutes \\
\midrule
 & &  & \textbf{Immune cSCC} & & \\
\midrule
ACTINN              & 0.9646 & 0.9315                       & 40,027             & 6,994                      & \textbf{2.4 Minutes} \\
N-ACT (Ours)        & \textbf{0.9684} & \textbf{0.9455}     & 40,027             & 6,994                      & 7.7 Minutes \\
\bottomrule
\end{tabular}
\end{sc}
\end{small}
\end{center}
\vskip -0.3in
\end{table}

\section{ Utility of N-ACT for Disambiguation of Broad Annotations}
\label{sec:appendix-azimuth}

 \begin{figure}[H]
    \centering
    \includegraphics[width=0.8\textwidth]{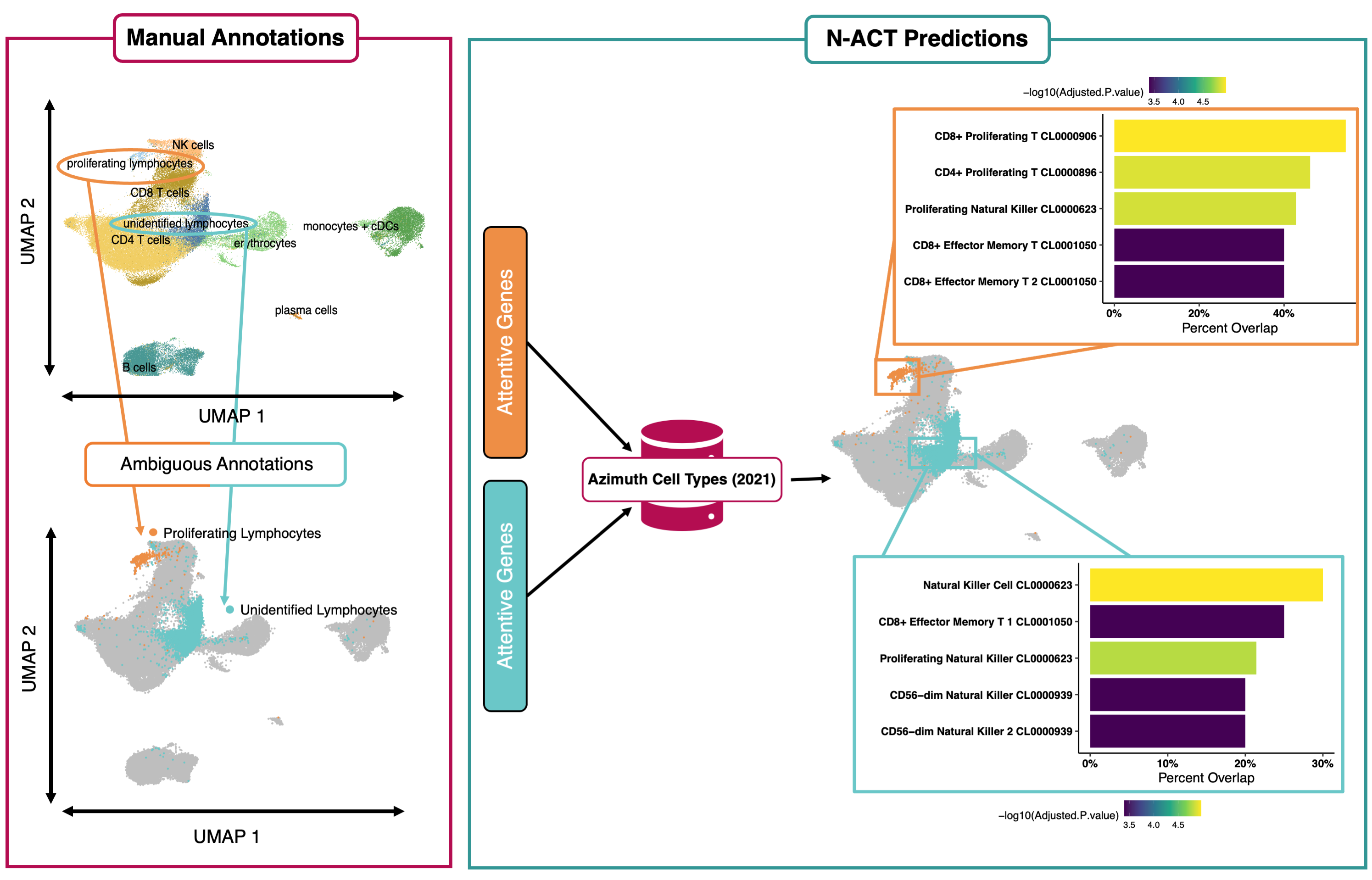}
    \smallskip
    \caption{\textbf{N-ACT Used to Disambiguate Broad Annotations}. Here, we use N-ACT-identified salient genes to query a specialized database (Azimuth) to disambiguate original broad manual annotations.}
    \label{fig:appendix-azimuth}
\end{figure}

Given that many manual annotations are typically performed with only a few genes (often two or three \citeA{BestPracticesTheis}), it is possible to have populations that are broad and ambiguous. This was the case with two populations of COVID PBMC data \citeA{COVIDPBMC}, namely the original annotations ``Proliferating Lymphocytes" and `` Unidentified Lymphocytes" (Fig. \ref{fig:appendix-azimuth}). Here we utilize N-ACT to disambiguate these broad annotations without re-clustering or further complex analyses.

In the results shown in the main manuscript, we utilized CellMesh \citeA{CellMesh}. However, using specialized databases could provide refined predictions. To demonstrate this, we investigated the two broad annotations in the COVID PBMC data and queried attentive genes from the Azimuth Cell Type 2021 database tailored for immune cells \citeA{Azimuth} [using Enrichr \texttt{R} Package (v 3.0) \citeA{Enrichr1, Enrichr2, Enrichr3}], to perform an enrichment analysis. We query our top 25 genes against the Azimuth Cell Types 2021 reference for the enrichment analysis and select the top 25 attentive genes (as described in \ref{sec:results-unsuperACTI} and \ref{sec:appendix-math}) for ``proliferating lymphocytes" and ``unidentified lymphocytes." 

Our enrichment analysis for ``proliferating lymphocytes" resulted in multiple significant (significant adjusted $p-$values) populations, with the most probable types being CD8+ proliferating T cells (shown in \ref{fig:appendix-azimuth}). Intuitively, these results are expected given the quantitative and qualitative similarity of this population with the CD8+ T cell population. It is important to note that the difference in gene overlap percentages of top three predictions, namely CD8 proliferating T, CD4 proliferating T, and proliferating natural killer populations is very small, which may explain why the original annotations were left broadly as proliferating lymphocytes. Enrichment analysis for ``unidentified lymphocytes" yielded natural killer cells as the most probable cell type, with other viable populations also being statistically significant. Similar to the previous population, our results are possibly intuitive given the closeness of the proliferating lymphocytes to CD8+ T cells and natural killer cells. Additionally, there is known overlap in the gene sets between of CD8+ T cells and natural killer cells. Lastly, we note that the second most probable population based on attentive genes are CD8+ effector memory T, suggesting that the ``unidentified lymphocyte" population could likely be refined into two or more populations. These results further signify the utility of N-ACT for unsupervised annotation, and the applicability of our framework in tandem with other annotation forms to provide interpretability and validation. 

\clearpage
\bibliographystyleA{StyleFiles/icml2022}
\bibliographyA{A}

\end{document}